\pdfoutput=1
\documentclass{aa}  

\usepackage{graphicx,natbib}
\usepackage{txfonts}
\begin{document}

   \title{Magnetically aligned dust and SiO maser polarization in the
     envelope of the red supergiant VY~CMa}

   \titlerunning{Magnetically aligned dust and SiO maser polarization around VY~CMa}


   \author{W.~H.~T. Vlemmings
          \inst{1}\fnmsep\thanks{wouter.vlemmings@chalmers.se}
          \and
          T. Khouri\inst{1} \and I. Mart{\'i}-Vidal\inst{1} \and
          D. Tafoya\inst{1} \and A. Baudry\inst{2} \and S. Etoka\inst{3} \and
          E.~M.~L. Humphreys\inst{4} \and T.~J. Jones\inst{5} \and
          A. Kemball\inst{6} \and E. O'Gorman\inst{7} \and
          A.~F. P{\'e}rez-S{\'a}nchez\inst{8} \and A.~M.~S. Richards\inst{9}
          }

   \institute{Department of Earth and Space Sciences, Chalmers University of Technology, Onsala Space Observatory, 439 92 Onsala, Sweden
          \and
           Laboratoire d'astrophysique de Bordeaux, Univ. Bordeaux, CNRS, B18N, all\'ee Geoffroy Saint-Hilaire, F-33615 Pessac, France
         \and
           Hamburger Sternwarte, Gojenbergsweg 112, 21029 Hamburg, Germany
         \and
           European Southern Observatory, Karl-Schwarzschild-Str. 2, 85748 Garching, Germany
        \and
          Minnesota Institute for Astrophysics, University of Minnesota, Minneapolis, MN 55455
       \and
          Department of Astronomy and National Center for
          Supercomputing Applications, University of Illinois at
          Urbana-Champaign, 1002 W. Green Street, Urbana, IL, 61801,
          USA
          \and
          Dublin Institute for Advanced Studies, 31 Fitzwilliam Place, Dublin 2, Ireland
          \and
          European Southern Observatory, Alonso de C\'ordova 3107, Vitacura,
          Casilla 19001, Santiago, Chile
          \and
          Jodrell Bank Centre for Astrophysics, School of Physics and Astronomy, University of Manchester, Manchester M13 9PL, UK
}

   \date{06-Mar-2017}

 
  \abstract
   {}
   {Polarization observations of circumstellar dust and molecular
     (thermal and maser) lines provide unique information about dust properties
     and magnetic fields in circumstellar envelopes of evolved
     stars.}
   {We use Atacama Large Millimeter/submillimeter Array Band 5 science
     verification observations of the red
     supergiant VY~CMa to study the polarization of SiO thermal/masers
     lines and dust continuum at $\sim1.7$~mm wavelength. We analyse
     both linear and circular polarization and derive the magnetic
     field strength and structure, assuming the polarization of the
     lines originates from the Zeeman effect, and that of the dust
     originates from aligned dust grains. We also discuss other
     effects that could give rise to the observed polarization.}
   {We detect, for the first time, significant polarization
     ($\sim3\%$) of the circumstellar dust emission at millimeter
     wavelengths. The polarization is 
     uniform with an electric vector position angle of $\sim8^\circ$. Varying levels of
     linear polarization are detected for the $J=4-3$~$^{28}$SiO~$v=0,~1,~2,$
     and $^{29}$SiO~$v=0,~1$ lines, with the strongest polarization
     fraction of $\sim30\%$ found for the $^{29}$SiO~$v=1$ maser. The
     linear polarization vectors rotate with velocity, consistent with
     earlier observations.  We also find significant (up to $\sim1\%$)
     circular polarization in several lines, consistent with previous
     measurements. We conclude that the detection is robust against
     calibration and regular instrumental errors, although we cannot
     yet fully rule out non-standard instrumental effects.}
   {Emission from magnetically aligned grains is the most likely
     origin of the observed continuum polarization. This implies that the dust is
     embedded in a magnetic field $>13$~mG. The
     maser line polarization traces the magnetic field
     structure. The magnetic field in the gas and dust is consistent
     with an approximately toroidal field configuration, but only
     higher angular resolution observations will be able to reveal
     more detailed field structure.  If the circular polarization is
     due to Zeeman splitting, it indicates a magnetic field strength of
     $\sim1-3$~Gauss, consistent with previous maser
     observations.}

   \keywords{supergiants, stars: individual: VY Canis Majoris, stars:
     mass-loss, magnetic fields, polarization, masers
               }

   \maketitle
%

\section{Introduction}

The study of the stellar winds of evolved massive stars is important
not only because the amount of expelled mass is a key factor in
determining the subsequent evolution of the star, and its eventual
fate, but also because they have a profound impact on the
interstellar medium and evolution of the Galaxy via their material and
energy output. However, the mechanisms that drive mass loss from these
massive stars are not well understood. It is generally assumed that
the acceleration of the stellar wind in low- to intermediate mass
asymptotic giant branch (AGB) stars is mainly due to radiation
pressure on dust grains \citep[e.g.][and references therein]
{Habing1996,Hoffner2008}. However, the formation of dust grains that
are non-transparent to stellar radiation is inhibited in the
vicinities of evolved massive stars due to their extreme luminosity
\citep{Speck2000, Woitke2006, Norris2012}. It is thus puzzling how the
material is transported from the surface of the star to the radius
where non-transparent dust species form. Furthermore, it has been
observed that the distribution of material in the circumstellar
envelope (CSE) of red supergiant stars is far from spherical
\citep[e.g.][]{Humphreys2007, OGorman2015}. This suggests the action
of an aspherical mechanism in the acceleration of the gas. While
several models have been proposed to explain the (anisotropic) mass
loss in evolve massive stars \citep[including convection, pulsation,
scattering off large grains, acoustic and magnetic waves;
e.g. ][]{Hartmann1980, Josselin2007, Hoffner2008}, it remains unclear
which of these mechanisms plays the dominant role.

Spectroscopic imaging of the outflows in evolved stars give important
clues about their geometry and velocity fields, thus the physics that
governs them. In addition, polarimetric observations of the molecular
and continuum emission provide valuable information on the magnetic
field and characteristics of the dust grains. Specifically,
polarimetric continuum observations can reveal the common orientation
of elongated dust grains around evolved stars due to the effect of a
global magnetic field in a similar way as has been observed in star
forming regions \citep[e.g.][]{Cortes2016}, and planetary nebulae
\citep[e.g.][]{Sabin2014}. Observations of polarized maser emission,
from molecules such as SiO, H$_{2}$O and OH, have also been important
in the determination of the strength and geometry of the magnetic
field in the expanding envelopes of evolved stars
\citep[e.g.][]{Kemball1997,Vlemmings2005, Etoka2010}. This type of
observations can be used to probe the magnetic field in the
circumstellar envelope (CSE) of evolved massive stars and determine
its relation to the mass loss phenomenon.
 
VY CMa is an excellent example of a massive evolved star with a high
mass loss rate ($\sim$$10^{-4} M_{\odot}$~yr$^{-1}$) and high dust
content in its CSE \citep{Danchi1994,Decin2006,Humphreys2007}, which
makes it an attractive target to explore the processes of mass loss
and dust formation in massive stars
\citep{Richards2014,DeBeck2015,OGorman2015}. It is one of the largest
red supergiant stars, located at a distance of 1.2 kpc
\citep{Zhang2012}. Recent high angular resolution ALMA Band 7 (around
$\sim321$~GHz) and 9 (around $658$~GHz) observations, tracing scales
down to $60$~mas, revealed the presence of two enigmatic features
detected only in continuum emission at submillimetre wavelengths. The
brightest of them, dubbed ``clump C'', is located 334 mas south-east
of VY CMa, while the other structure, VY, is located toward the north,
elongated in the direction perpendicular to the line that connects the
star and clump C \citep[][hereafter OG+15]{OGorman2015}. It is
unlikely that these structures formed from convection cells since it
would require a constant mass loss in a single direction for a few
decades, but it has been suggested that they could have been formed as
a result of localised long-lived MHD disturbances in the photosphere
(OG+15). From polarization observations of the maser emission it has
been found that the strength of the magnetic field in the CSE ranges
from few mG to few G
\citep{Vlemmings2002,Herpin2006,Richter2016}. Such observations
however, do not clearly reveal the global structure of the magnetic
field within the CSE. In this paper we present the analysis of ALMA
full polarization Band 5 science verification observations toward the
star VY~CMa and report the detection of polarized line and continuum
emission around the star that is used to probe the magnetic field and
determine its association with features seen in its CSE.

 \begin{figure*}
 \centering
 \includegraphics[width=15cm]{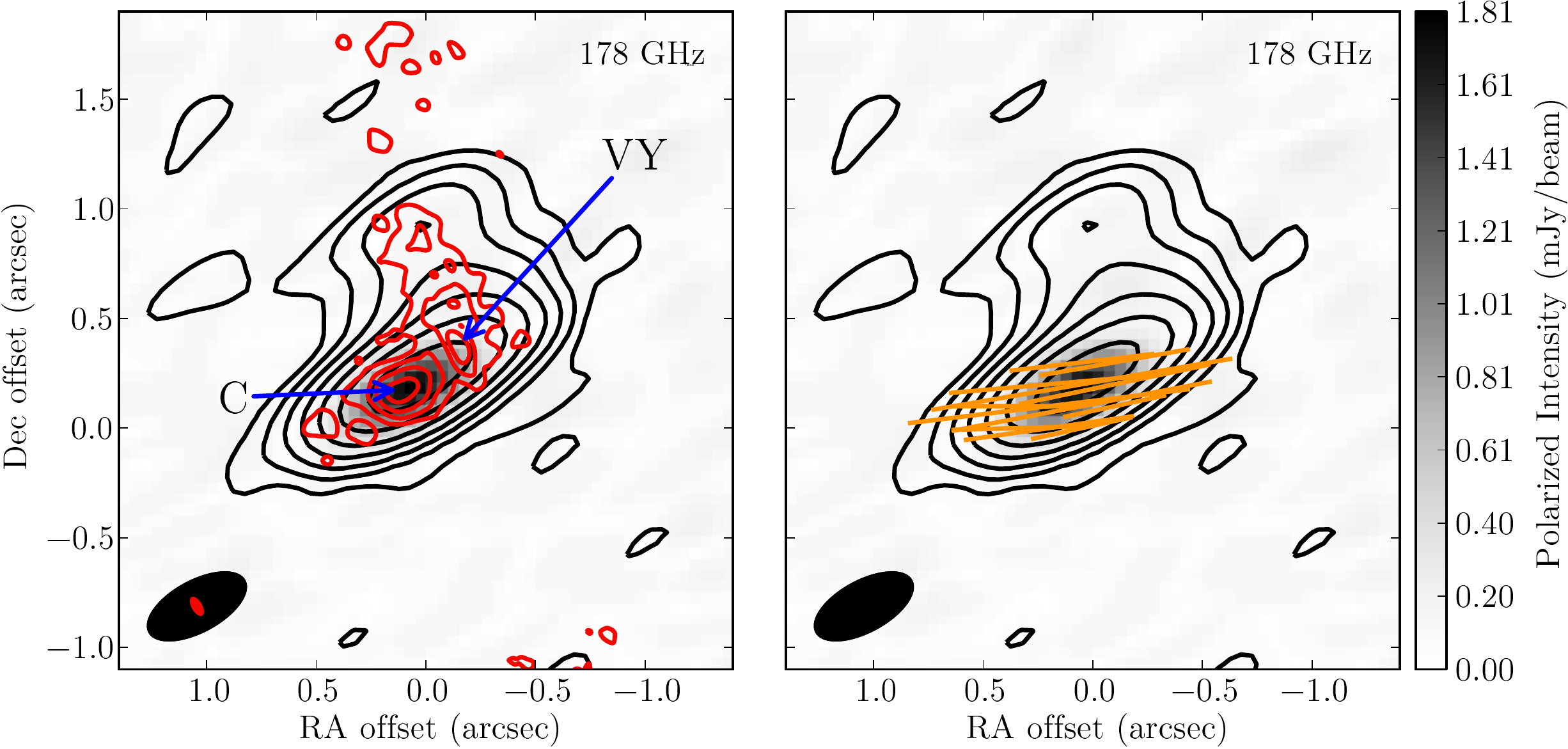}
 \caption{The dust at 178~GHz around of VY CMa. Arrows indicate dust clump C
   and the star (VY) identified in previous ALMA observations \citep[][OG+15]{Richards2014}. The grey scale image
   is the linearly polarized intensity, which is seen to peak at the
   bright dust component Clump C in the South-East. The black contours
   are the total intensity at $0.01, 0.02, 0.04, 0.08, 0.16, 0.32$,
   and $0.64$ times the peak brightness. The similarly spaced red contours
   (left) indicate the ALMA 658~GHz continuum from OG+15. The vectors
   (right) indicate the direction of polarization rotated by
   $90^\circ$ to indicate the magnetic field direction if the
   polarized emission originates from magnetically aligned grains. The
   beam size is indicated in the lower left corner.}
       \label{cont}
 \end{figure*}

\section{Observations and data reduction}

\begin{table*}
\caption{Spectral window setup}
\centering
\begin{tabular}{ c c c c c c c}
\hline \hline
Spw & Center frequency & Channel Width & & Number of Channels & Total
Bandwidth & Line(s) \\ 
 & [GHz] & [MHz] & [km~s$^{-1}$]& & [GHz] & \\
\hline 
0 & 183.310 & 0.244 & 0.40 & 960 & 0.2344 & para-H$_2$O $3_{13}-2_{20}$\\ 
1 & 183.884 & 1.953 & 3.18 & 960 & 1.8750 & -\\ 
2 & 170.999 & 1.953 & 3.42 & 960 & 1.8750 & $^{28}$SiO $(J=4-3, v=2)$\\
  &                &           &         &        &             & $^{29}$SiO $(J=4-3, v=0)$\\  
  &                &           &         &        &             & $^{29}$SiO $(J=4-3, v=1)$\\  
3 & 172.481 & 0.244 & 0.42 & 480 & 0.1172 & $^{28}$SiO $(J=4-3, v=1)$\\ 
4 & 173.668 & 0.244 & 0.42 & 480 & 0.1172 & $^{28}$SiO $(J=4-3, v=0)$\\ 
\hline 
\end{tabular} 
\label{SPWTAB}
\end{table*}

These observations are part of the ALMA Band 5 ($163-211$~GHz) Science Verification
(SV) campaign. The VY\,CMa observations took place on 16 October 2016,
starting at 05:45\,UT and lasting for about 6.4\,hours. A total of 13
antennas were used in the calibration and imaging.  The observations
were performed using five spectral windows (spws). The spw setup is
presented in Table.~\ref{SPWTAB}.

The full calibration and imaging was performed in CASA 4.7.0, and is
described in the Band 5 polarization calibration memo \citep{IMV2016}
that was released with the data. The gain calibration (bandpass,
amplitude, absolute flux-density scale and phase) was performed
following the standard ALMA procedures. The amplitude and phase gains
for spw\,0 (i.e., the window centered at 183.299\,GHz), which has the
narrowest bandwidth (234.4\,MHz), were derived by transferring the
gain solutions from spw\,1 (total bandwidth of 1.875\,GHz).

The polarization calibration was also performed following standard
ALMA procedures, although with some improvements that we summarize
here. The X-Y cross-polarization phases at the
reference antenna were fitted by pre-averaging spw\,0 and spw\,1 in
chunks of 200 and 20 channels, respectively. This was done to increase
the SNR of the gain solutions close to the atmospheric water-vapour
line. Also, the calibration tables were created on a
per-spw basis. This was needed to properly derive the ambiguities of
the X-Y phases (see \cite{IMV2016}, Sect. 4.2, for more details).

Finally, we performed four iterations of self-calibration on a strong channel of the SiO
maser in spw~3. These solutions were applied to all the other spw.  We
then reimaged all the spws at full spectral resolution and both the
continuum and lines at full polarization, producing data cubes in
Stokes I, Q, U, and V. The imaging of the spectral lines was performed
using Briggs weighting with a robustness parameter of $0.5$, which
resulted in a synthesized beam of $\sim0.8\times0.3\arcsec$. The
typical rms noise in line free channels around the SiO lines was
$\sim3.8$~mJy~beam$^{-1}$ for spw~3, and 4, and
$\sim1.4$~mJy~beam$^{-1}$ for the coarser spectral resolution of
spw~2. The continuum was imaged using a robust gridding weighting
parameter of $-0.5$ resulting in
a beam of $\sim0.5\times0.2\arcsec$. We achieved an rms noise of
$0.1$~mJy~beam$^{-1}$.

According to the calibration procedure described in \cite{IMV2016} and
used here, the estimates of the instrumental polarization in the spws
close to the atmospheric water line (i.e., spw\,0 and spw\,1) may be
biased due to the higher noise in the water-vapour frequency window
and/or to the mapping function (i.e., dependence with antenna
elevation) of the atmospheric absorption. The polarization calibration
in these spws should thus be taken with care. With respect to
spw\,2$-$4, the polarization calibration is not affected by such a
bias. Therefore, we use only the linear-polarization results from
spw\,2$-$4 (covering the SiO masers and the continuum) in our
analysis. Regarding the images in circular polarization (i.e., Stokes
V), a Stokes V signal can be affected by defects or biases in the
calibration of the instrumental polarization; from a biased estimate
of the phase between the polarizers at the reference antenna to biased
(and correlated) estimates of the polarization leakage at all the
antennas. A justification of the fidelity of these Stokes V images,
together with a robustness test of the instrumental polarization
calibration of these data, is described in Appendix
\ref{AppMC}). Based on this analysis, we rule out calibration errors
or standard instrumental effects as the origin of the observed
circular polarization. However, as these are the first reported ALMA
circular polarization results, there is still the chance that they are
affected by unkown instrumental issues.

\section{Results}

 \begin{figure*}
\centering
\includegraphics[width=15cm]{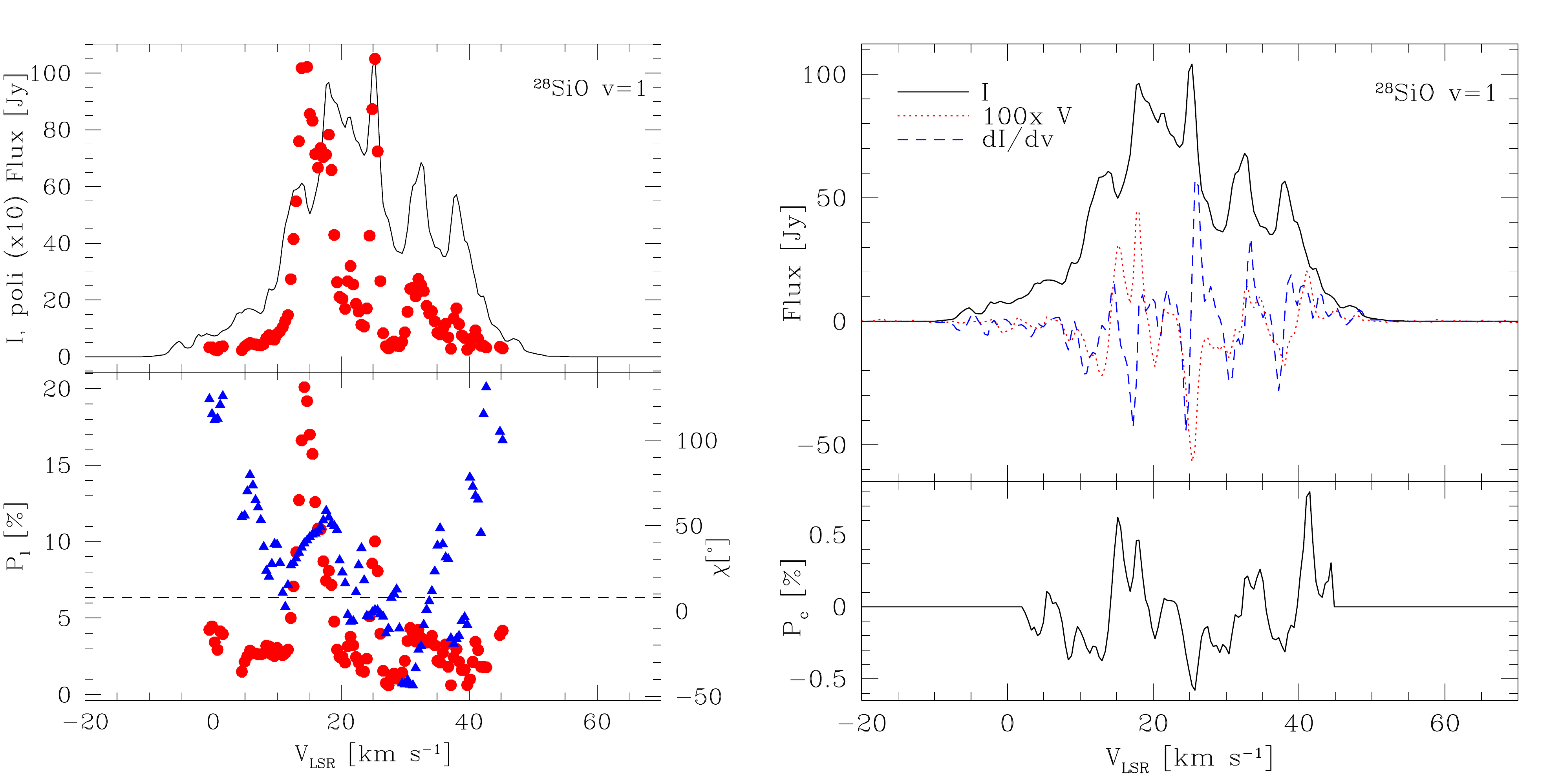}
\caption{{\it (left, top)} Total intensity (solid line) and linearly
  polarized intensity (solid red circles) from
 the $^{28}$SiO ($J=4-3$, $v=1$) transition. The linear polarization
 spectrum is multiplied by a factor of 10. {\it (left, bottom)} The
 linear polarization fraction (red dots) and electric vector position angle (EVPA) (blue
 triangles). The horizontal dashed line indicates the EVPA observed in the dust continuum. {\it (right, top)} The
 circularly polarized emission (red short-dashed, multiplied by 100)
 for the $^{28}$SiO ($J=4-3$, $v=1$) transition. Overplotted is also
 the total intensity derivative $dI/d\nu$ (blue long-dashed) scaled
 for a magnetic field strength of $\sim2.4$~G.{\it (right, bottom) }
 The circular polarization fraction $P_c$, which is found to peak at
 $\sim0.8\%$ for this transition.  }
    \label{line1}
\end{figure*}

\begin{figure*}
\centering
\includegraphics[width=15cm]{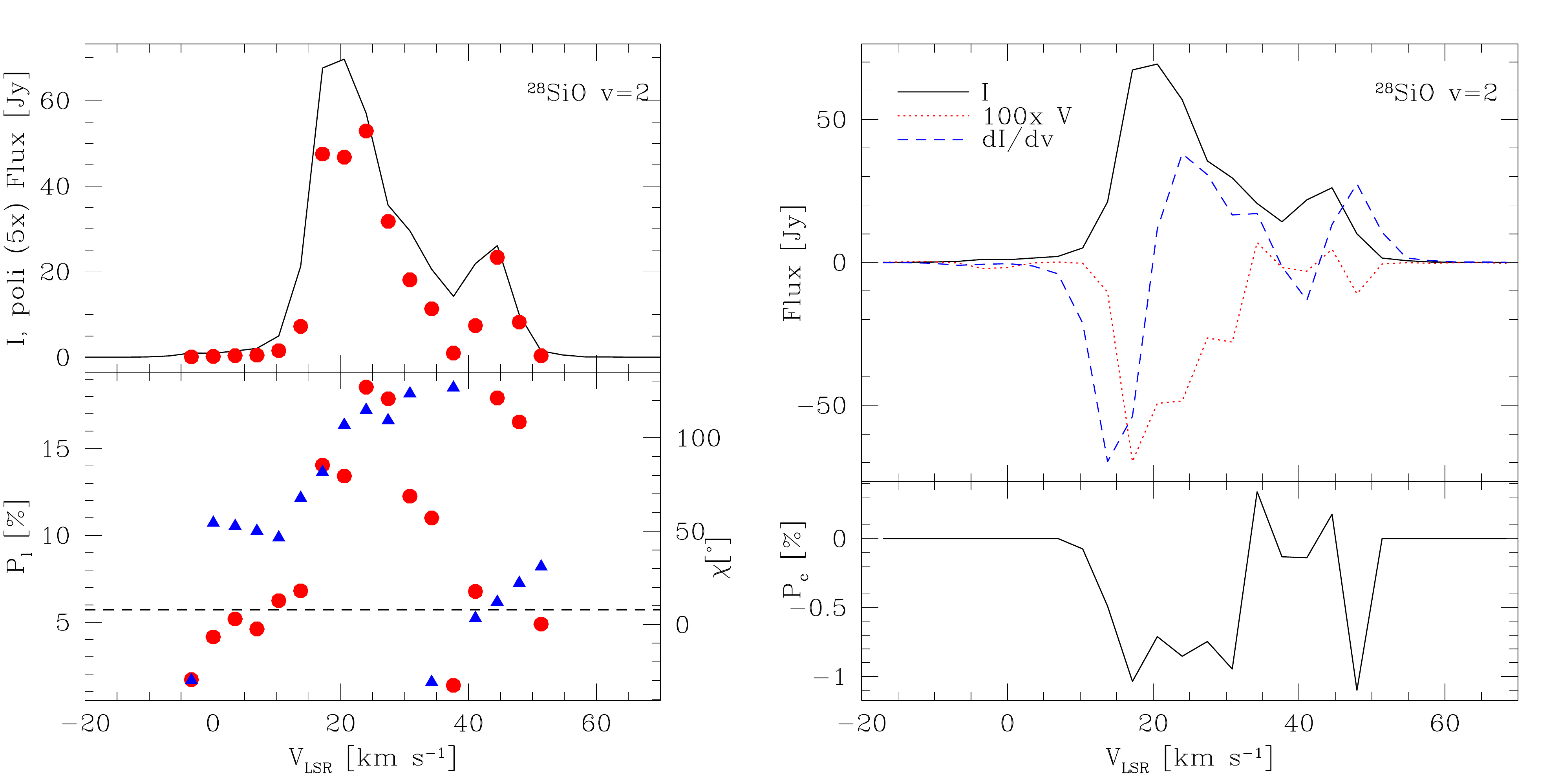}
 \caption{As Fig.~\ref{line1} for the $^{28}$SiO ($J=4-3$, $v=2$)
   transition. The linear polarization spectrum is multiplied by five.
}
    \label{line2}
\end{figure*}
\begin{figure*}
\centering
\includegraphics[width=15cm]{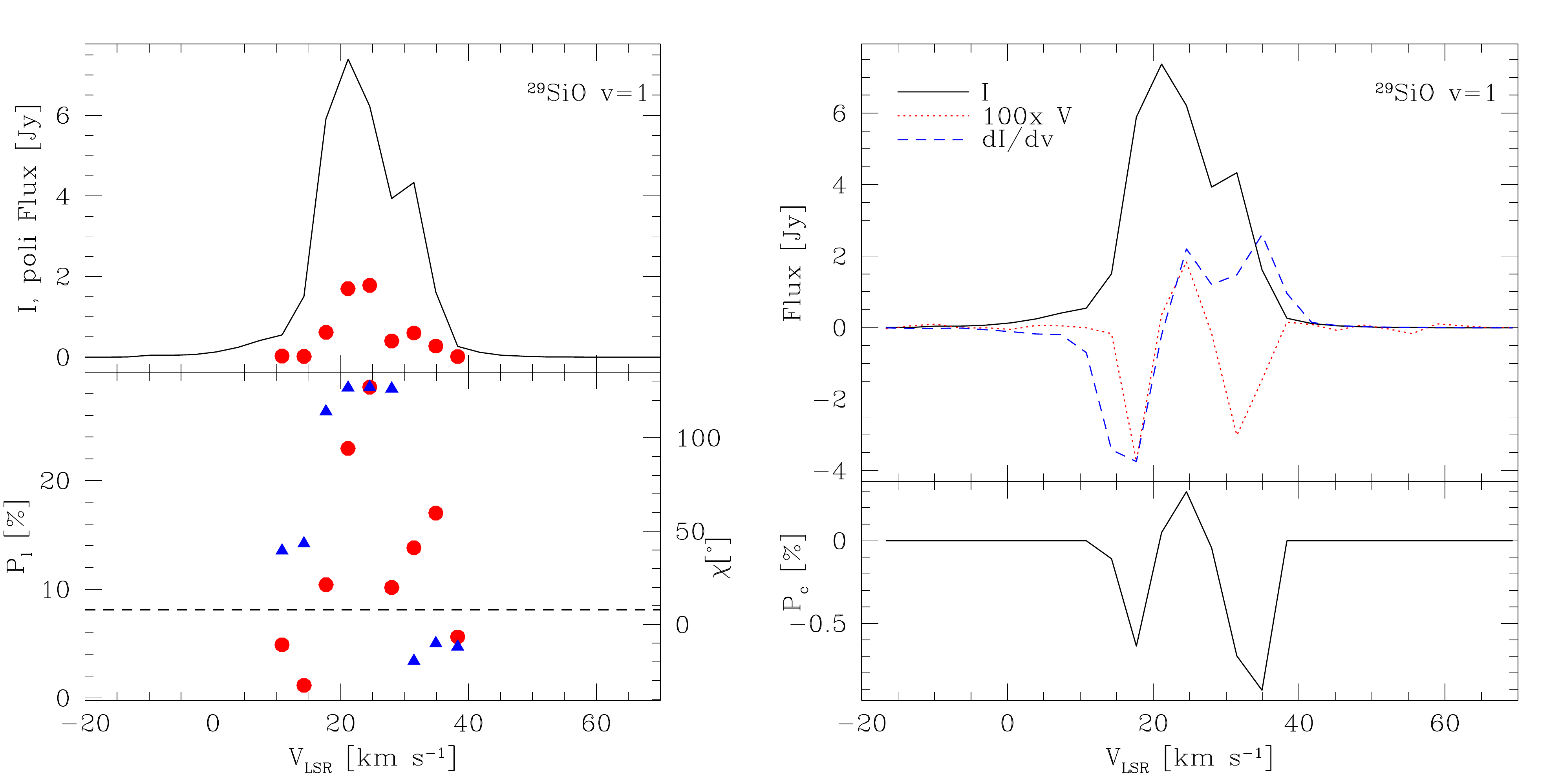}
 \caption{As Fig.~\ref{line1} for the $^{29}$SiO ($J=4-3$, $v=1$)
   transition. The linear polarization spectrum is not multiplied.
}
    \label{line3}
\end{figure*}

\subsection{Dust continuum}

In order to map the continuum, we only used line-free channels
sufficiently far from the atmospheric H$_2$O line. The resulting
aggregate bandwidth was 2.25~GHz. The VY~CMa continuum, centered
around 178~GHz, is extended. Although the beam is elongated exactly
along the direction between component VY and the continuum clump C
seen in (OG+15), the distribution appears identical when the ALMA Band
7 and Band 9 observations (OG+15) are convolved with the Band 5
beam. In Fig.~\ref{cont}(left) we show the Band 5 continuum
overplotted on the ALMA Band 9 observations around
658~GHz. Deconvolving the two components, we measure a total flux of
$36\pm3$~mJy for the star and $94\pm9$~mJy for clump C. These values
are completely consistent with the average spectral index of the two
component determined from the Band 7 ($321$~GHz) and Band 9 ($658$~GHz) observations
(OG+15). This implies that clump C is still optically thick at 178~GHz
and increases, following the analysis from OG+15, the dust mass to
$>1.2\times10^{-3}$~M$_\odot$ and decreases the dust temperature to
$\sim100$~K. As in OG+15, we did not detect the SW clump seen in
Hubble observations of scattered light and a number of molecular lines
\citep[e.g.][]{Humphreys2007, DeBeck2015}.

We also detect clear linear polarization. The maximum fractional
polarization is $\sim3\%$. The polarized intensity is indicated in
greyscale in Fig.~\ref{cont} and clearly peaks at the location of
Clump C. Despite the large beam, there is no polarization found
associated with VY or the North-East dust extension, even though
the sensitivity would have been sufficient to detect both down to
$1-2\%$. In Fig.~\ref{cont}(right) we show the polarization vectors
rotated by $90^\circ$ to reflect the magnetic field direction if the
polarization is due to the emission from magnetically aligned
grains. The average (non-rotated) electric vector position angle (EVPA) is $8\pm4^\circ$
(East of North),
where the error includes both the scatter around the average and the
error due to the noise in the polarization images. The vectors are
thus neither tangential nor parallel to the direction of the star to
clump C, which has a position angle of $\sim130^\circ$.

\subsection{SiO emission}

\begin{table*}
\caption{Observed SiO lines}             
\label{tab1}      
\centering          
\begin{tabular}{l l c c c c c }     
\hline\hline       
Rest Frequency &  Transition    & E$_u$/k & $A$ & Peak Flux & $P_{l,{\rm max}}$&
$|P_{c,{\rm max}}|$ \\
$[{\rm GHz}]$ &   &  [K] & [s$^{-1}$] & [Jy] & [$\%$] & [$\%$] \\
\hline
\multicolumn{6}{c}{$^{28}$SiO} \\
\hline
173.688\tablefootmark{a}     & $J=4-3, v=0$ &  21  &  $2.602\times 10^{-4}$ & 2.4 &  4.0 &  1.4 \\ 
172.481\tablefootmark{a}     & $J=4-3, v=1$ & 1790 & $2.569\times
10^{-4}$&  104.2 &  20.1 &   0.8 \\ 
171.275\tablefootmark{b}  & $J=4-3, v=2$ & 3542  & $2.547\times 10^{-4}$
&  69.4 &  18.5 & -1.1  \\ 
\hline
\multicolumn{6}{c}{$^{29}$SiO}\\
\hline
171.512\tablefootmark{b}     & $J=4-3, v=0$ &  21 &  $2.505\times
10^{-4}$  &  3.2 &  2.0 &  <0.3 \\
170.328\tablefootmark{b}     & $J=4-3, v=1$ &  1779 &  $2.484\times
10^{-4}$ &  7.4 &  28.6 &  0.9  \\      
\hline                  
\end{tabular}
\tablefoot{
\tablefoottext{a}{Observed at $0.42$~km~s$^{-1}$ spectral resolution.}
\tablefoottext{b}{Observed at $3.4$~km~s$^{-1}$ spectral resolution.}}
\end{table*}

\begin{figure*}
\centering
\includegraphics[width=15cm]{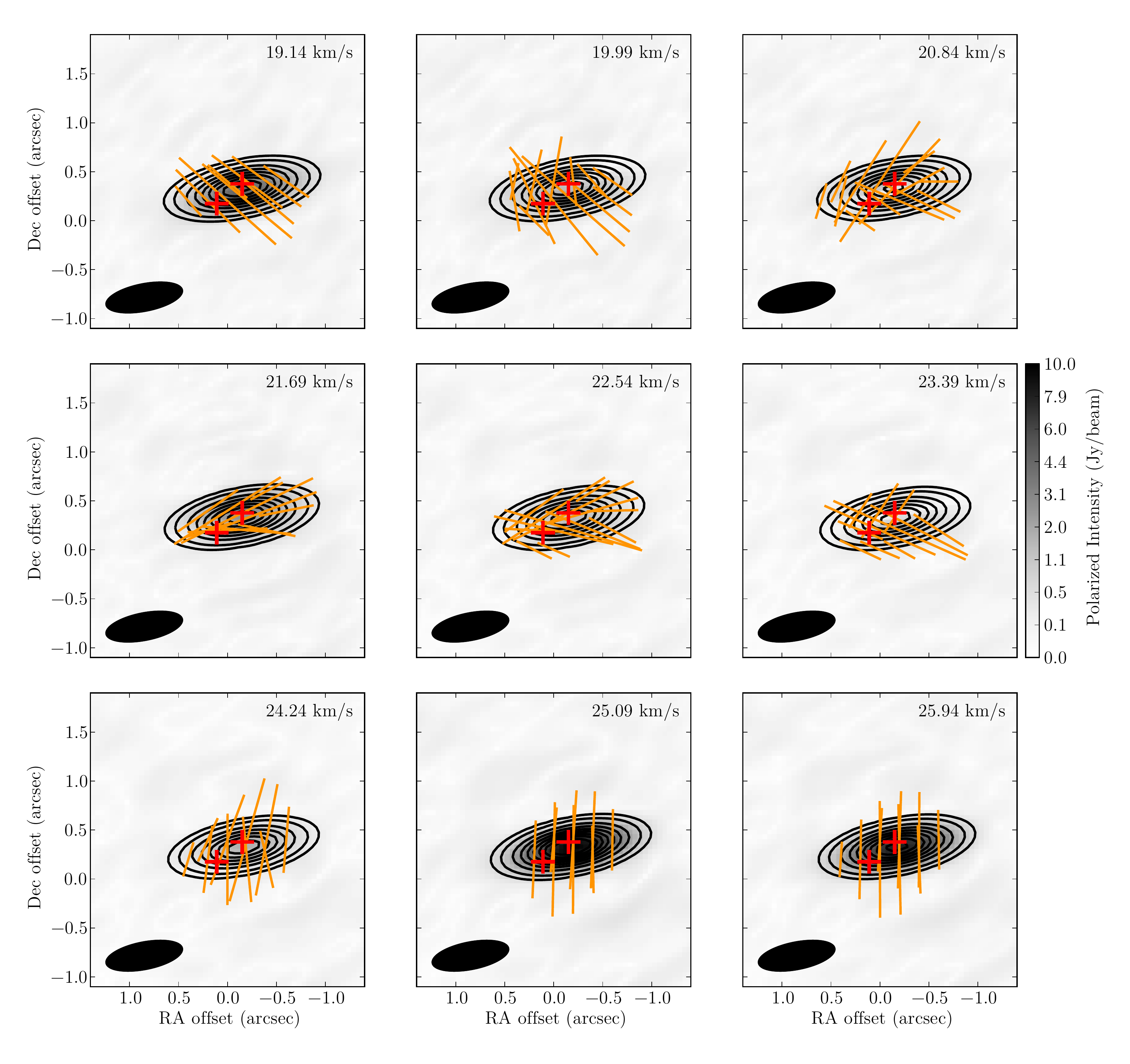}
 \caption{Channel maps of the linear polarization of the $^{28}$SiO
   ($J=4-3$, $v=1$) maser at select channels around the stellar velocity
   ($V_{\rm LSR}=22$~km~s$^{-1}$). The (red) crosses indicate the
   position of the star (close to the centre of the image) and clump C
   (towards the south-east). The grey scale image is the linearly
   polarized intensity. The contours are the total intensity
   at $0.05, 0.1, 0.2, 0.3, 0.4, 0.5, 0.6, 0.7$, and $0.8$ times the
   peak flux. The vectors indicate the direction of polarization. The beam
   size is indicated in the lower left corner.}
    \label{SiOv1}
\end{figure*}

The observational setup included several lines of SiO and its
$^{29}$SiO isotope. They display maser emission in the vibrationally
excited states and predominantly thermal emission in the $v=0$
state. These SiO maser and thermal emissions are associated with the
VY component towards the star, and not with clump C. The line strength
and polarization of the SiO lines covered are listed in
Table~\ref{tab1} along with their transition energy and Einstein $A$
coefficients \citep{Muller2001}. In the rest of the paper we omit
$J=4-3$ when we describe the transitions. All the maser lines show a
complicated line structure, but are not, or only marginally, spatially
resolved in our $0.8\times0.3\arcsec$ beam. Their spectra, in Stokes
I, linear polarization fraction, EVPA, and Stokes V, are
shown in Figs.~\ref{line1},~\ref{line2}, and~\ref{line3}.  The
strongest linear polarization, with a fraction up to $\sim30\%$, is
detected for the $^{29}$SiO $v=1$ transition with polarization up to
$\sim20\%$ detected for $^{28}$SiO $v=2$. With the exception of a
single, possibly saturated maser feature (as discussed in
section~\ref{SiOpol}), the polarization of the $^{28}$SiO $v=1$ line
is around $5\%$. The linear polarization fraction is consistent with
estimates based on recent APEX observations of the same transitions
\citep{Humphreys2017}. In Fig.~\ref{SiOv1}, we present a number of
channel maps of the linear polarization for this transition. For all
maser lines we also detect cicular polarization at a level up to
$\sim1\%$. The spectral resolution for all but the $^{28}$SiO $v=1$
line is not sufficient for a more detailed analysis. Hence, unless
otherwise noted, the rest of our analysis is based on the $^{28}$SiO
$v=1$ transition.

 \begin{figure*}
\centering
\includegraphics[width=15cm]{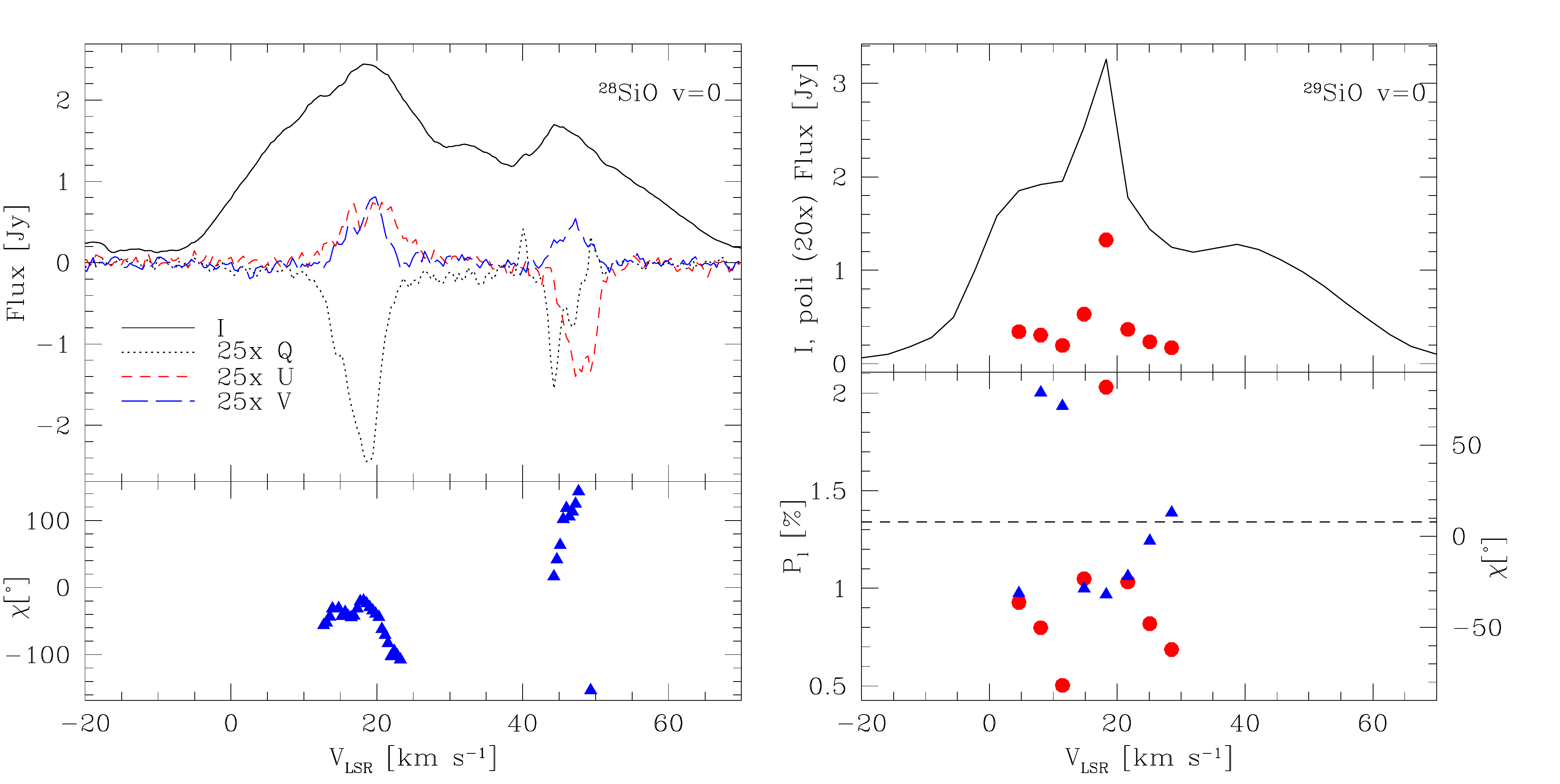}
  \caption{{\it (left)} The Stokes I, Q, U, and V spectra for the
    $^{28}$SiO ($J=4-3$, $v=0$) transition extracted in the region where
    linear polarization was detected. The bottom panel displays the
    EVPA (blue triangles) determined from the Q, and U
    spectra. {\it (right)}  As Fig.\ref{line1}(left) for the $^{29}$SiO ($J=4-3$, $v=0$)
    transition. The linear polarization spectrum is multiplied by
    20.
}
     \label{SiOv0spec}
\end{figure*}


\begin{figure*}
\centering
\includegraphics[width=15cm]{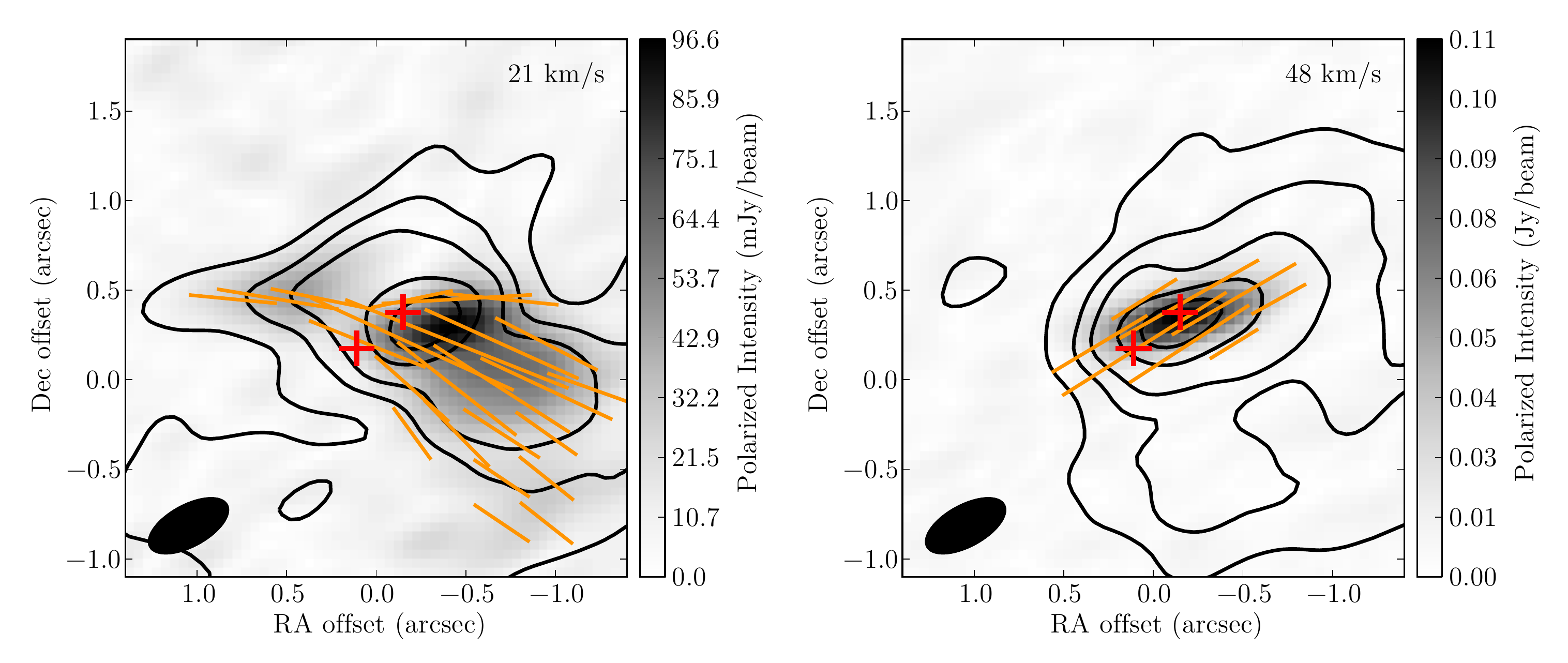}
  \caption{Channel images of the thermal $^{28}$SiO ($J=4-3$, $v=0$)
    polarization of the two polarized components seen in
    Fig.~\ref{SiOv0spec}. The crosses, grayscale, contours and vectors are
    as Fig.~\ref{SiOv1} with contours at $0.2, 0.3, 0.4, 0.6$, and
    $0.8$ times the peak.}
     \label{SiOv0}
\end{figure*}
 
The thermal SiO $v=0$ line is significantly spatially extended, both
in Stokes I as well as in linear polarization. The polarization
spectrum in Fig.~\ref{SiOv0spec}, extracted in an apperture
encompassing the extended polarized emission, shows two distinct
spectral features. For illustration, two channel maps representing the
two components, blue-shifted at $V_{\rm LSR}=21$~km~s$^{-1}$ and
red-shifted at $V_{\rm LSR}=48$~km~s$^{-1}$, are shown in
Fig.~\ref{SiOv0}. The neighbouring channels show a similar
structure. The polarization pattern at $V_{\rm LSR}=21$~km~s$^{-1}$
shows that the fractional polarization peaks away from VY and that the
polarization vectors point radially towards the star. However, at
$V_{\rm LSR}=48$~km~s$^{-1}$, the polarization peaks at the stellar
position and does not show the radial pattern. The two linearly
polarized components are also seen in the circular polarization
spectrum, which shows no sign of a typical S-shaped Zeeman pattern.

\section{Discussion}

\subsection{SiO maser polarization}
\label{SiOpol}

In the framework of basic maser theory, it is important to investigate
the rate of stimulated emission of the maser $R$, the maser decay rate
$\Gamma$ and the Zeeman frequency $g\Omega$, in order to determine if
the SiO maser polarization is the result of Zeeman or non-Zeeman
properties \citep[][and references therein]{PerezSanchez2013}. If
$g\Omega>R$, and the maser is not strongly saturated ($R\lesssim10\Gamma$), the
magnetic field determines the linear and circular polarization of the
maser. Under those conditions, the maximum fractional linear polarization
for the SiO transitions we observed will be of the order of $15\%$,
depending on the angle $\theta$ between the maser propagation
direction and the magnetic field. However, SiO masers often display
higher levels of fractional polarization. This is likely due to
anisotropic pumping of the energy levels involved in the maser
\citep[e.g.][]{Nedoluha1990}. However, even in that case, the
relationships between $R$, $\Gamma$, and $g\Omega$ still determine whether
the magnetic field can be traced by the maser polarization
\citep{Watson2002}.

Using the parameters from \citet{PerezSanchez2013},
$g\Omega\sim150$~s$^{-1}$ for the magnetic field strength of
$\sim0.1$~G in the H$_2$O maser region
\citep{Vlemmings2002}. The radiative decay rate $\Gamma$ can be
approximated by $\Gamma=5$~s$^{-1}$, which corresponds to the rate for
the radiative decay from the first vibrationally excited state to the
ground-vibrational state \citep{Nedoluha1990}. The stimulated emission
rate $R$ depends on the maser brightness temperature $T_b$, beaming
angle $\Delta\Omega$, Einstein $A$ coefficient and transition
frequency $\nu$ through:
\begin{equation}
R\sim{{A k T_b \Delta\Omega}\over{4 \pi h \nu}}.
\end{equation} 
We adopt a value of $\Delta\Omega=10^{-2}$~sr. The
brightness temperature itself depends on the size of the maser
emission region. Based on VLBI observations \citep[e.g.][]{Richter2016},
we adopt a typical maser spot size of $\sim1$~mas. As VLBI
observations generally do not recover all maser flux observed with
larger angular resolution, this size corresponds to the core of the
maser emission and likely results in an overestimate of $T_b$. From
this, we find that $T_b<2\times10^9$~K for the $^{28}$SiO and
$T_b<1\times10^8$~K for the $^{29}$SiO masers. For the stimulated
emission rate, we thus find $R<50$~s$^{-1}$ for $^{28}$SiO, and
$R<8$~s$^{-1}$ for $^{29}$SiO. Thus, while the strongest of the
$^{28}$SiO masers could be approaching saturation, $g\Omega>R$ and
$g\Omega>\Gamma$. We thus expect the magnetic field to determine the
polarization characteristics of the majority of the maser
features. This agrees well with the observed fractional linear
polarization, which is mostly $<15\%$. Only one of the strongest maser
features of the $^{28}$SiO $v=1$ transition reaches a polarization
fraction of $\sim20\%$, which could imply that this feature is
strongly saturated. This feature, around $V_{\rm lsr}=15$~km~s$^{-1}$
also displays a change in polarization vector that could be the result
of saturation.

As shown in Fig.~\ref{line1}(right), the observations of the
$^{28}$SiO $v=1$ also reveal significant circular
polarization. Possible spurious circular polarization is discussed in
Appendix~\ref{AppMC}. Assuming a general Zeeman splitting origin, the
circular polarization spectrum $V$ is expected to be proportional to
the derivative of the total intensity $I$, with a proportionality
coefficient that
depends on the magnetic field strength. This assumes a constant
magnetic field strength with uniform direction throughout the
maser. As indicated by the linear polarization, such a uniform
structure is unlikely. Still, we show that $V\propto dI/d\nu$ is a
reasonable approximation of the observed spectrum if one assumes a
magnetic field reversal at $V_{\rm lsr}\sim15$ and $30$~km~s$^{-1}$,
which is close to symmetric around the stellar velocity ($V_{\rm
  lsr}\sim22$~km~s$^{-1}$). This strongly supports the basic Zeeman
interpretation of the circular polarization. Although the spectral
resolution was significantly worse, circular polarization was also
detected for the $^{28}$SiO $v=2$ and the $^{29}$SiO $v=1$ maser. Also
in this case $V\propto dI/d\nu$ is a good approximation. In order to
obtain an estimate of the magnetic field strength required to produce
the observed circular polarization, we use the fractional circular
polarization \citep[following][]{Herpin2006, Kemball1997,
  Elitzur1996}. For the $J=4-3$ masers, we find $B\approx1.3 P_c
\cos{\theta}$ assuming a typical maser linewidth of
$\sim1$~km~s$^{-1}$. We thus find $B/\cos{\theta} \approx
0.85$~G. This is similar to the field strengths of $1-2$~G found using
the same method from previous $J=1-0$ and $J=2-1$ SiO maser
observations \citep{Herpin2006, Richter2016}, supporting the Zeeman
interpretation of the observed circular polarization. The above
approximation assumes saturated masers \citep{Elitzur1996} and
slightly different results are obtained when applying the basic Zeeman
fitting that might be more applicable to the non-saturated masers we
observe. We thus also determine the field strength using
\begin{equation}
V=zB\cos{\theta}{{dI}\over{d\nu}}.
\end{equation}
Here, the proportionality term
$z=g\Omega\approx1500$~s$^{-1}$~[Gauss]$^{-1}$. Our fit shown in
Fig.~\ref{line1}(right) then yields a magnetic field strength of
$B\cos{\theta}\sim2.4$~G. Thus, in either method, the magnetic field
is indeed strong enough for $g\Omega>R$. Both field strength estimates
are easily sufficient to magnetically align the dust grains in clump C
in sufficiently short time scale (as described in \S~\ref{dust}).

Alternatively, \citet{Houde2014} suggests the SiO maser circular
polarization could arise from resonant scattering of magnetised foreground
SiO gas. This causes linear polarization to be converted to circular
polarization. The effect has been able to explain the circular
polarization of 43~GHz SiO masers. The conversion phenomenon depends on the
excitation of the molecular transition in the foreground, and the much
lower column density of $^{29}$SiO would lead to a conversion
efficiency lower than that of $^{28}$SiO. As we observe a similar
circular polarization fraction, it appears that the resonant
scattering does not contribute significantly to the maser transitions
presented here.

\subsection{Thermal SiO polarization}

Linearly polarized line emission from non-maser molecular lines can
arise due to differences in excitation of the magnetic sublevels
within a transition. This can occur already in a weak magnetic field
when the magnetic sublevels of a rotational state experience
anisotropic emission. This effect is generally called the
Goldreich-Kylafis effect (hereafter GK-effect) and is described in
\citet{Goldreich1981,Goldreich1982}. Alternatively, linear polarization in
a circumstellar envelope can occur from molecules with a preferred
rotation axis because of strong, radial, infrared emission from a
central star \citep{Morris1985}. Significantly polarized (thermal)
molecular line emission around evolved stars has so far been observed
for SiO $v=0$ and CO in the envelope of the M-type AGB star IK~Tau
\citep{Vlemmings2012} and for CS, SiS and CO around the C-type AGB star
IRC+10216 \citep{Glenn1997, Girart2012}. In the case of radiative molecular
alignment in the absence of a magnetic field, the linear polarization
should mainly be perpendicular to the radial direction. Already for
few $\mu$G magnetic field strength however, the molecular alignment in
much of the circumstellar envelope is determined by magnetic field
\citep{Morris1985}. In case of the GK-effect, the linear polarization is
along the magnetic field as long as the Zeeman splitting dominates the
collisional and spontaneous emission rates \citep{Kylafis1983}.

The behaviour of thermal line polarization will thus change depending
on optical depth, distance from the star, and magnetic field strength
and geometry. Our observations seem to indicate such a change, as
shown in Fig.~\ref{SiOv0spec}(right) and Fig.~\ref{SiOv0}. The
polarization is confined to a velocity range between $V_{\rm
  lsr}=11-25$~km~s$^{-1}$ close to and slightly blue-shifted from the
stellar velocity, and a clearly redshifted component between $V_{\rm
  lsr}=42-52$~km~s$^{-1}$. While the (slightly) blue-shifted component
displays a radial polarization pattern that could indicate that the
radiation field dominates the molecular alignment, the red-shifted
component is neither radial nor tangential. The red-shifted
polarization component also peaks towards the star, contrary to
prediction of radiative alignment. It thus appears that the
red-shifted component traces the magnetic field according to the
GK-effect. This implies a magnetic field position angle of $\sim40^\circ$ in
the most red-shifted part of the circumstellar envelope which is
consistent with the extreme velocity SiO masers.

We also detect significant circular polarization in the $^{28}$SiO v=0
thermal line. As seen in Fig.~\ref{SiOv0spec}(left), the circular
polarization spectrum is significantly different from that observed
for the maser lines. It displays similarities to the linear
polarization (Stokes Q, and U) spectra, although it does not show the
negatives seen in those. It peaks at approximately $1.6\%$ but
does not show a typical Zeeman spectrum. In Appendix~\ref{AppMC}, we
rule out calibration errors, although unknown instrumental effects in
the ALMA system can only be ruled out once the observations are
repeated. Non-Zeeman circular polarization has previously been
reported for CO and has been suggested to be due to the resonant
scattering of foreground molecules via the same mechanism suggested
for SiO masers \citep{Houde2014}. The excitations conditions of the
thermal SiO foreground gas can explain why the effect is seen in the
$v=0$ transition and not the maser transitions. The lower abundance of
$^{29}$SiO is then the reason why no circular polarization was
detected for the $^{29}$SiO $v=0$ line despite the total intensity being
of equal strength to the $^{28}$SiO $v=0$ line. 

Alternatively the velocity ranges displaying polarized emission could
be those including maser components. In this case, the circular
polarization of $^{28}$SiO $v=0$ may instead arise from weaklu
saturated maser features.  This can only be confirmed with future,
higher angular resolution observations.

\subsection{Dust polarization of clump C}
\label{dust}

We consider two possible scenarios to explain the observed polarized
radiation in the continuum emission towards clump C, scattering off
large dust particles or emission from non-spherical grains with a
preferred orientation direction. In our analysis, we also include that
observations of polarized visible light obtained with Hubble
\citep{Jones2007} and SPHERE/VLT \citep{Scicluna2015} show that the
polarization vectors seen towards clump C in visible wavelengths are
roughly perpendicular to those observed at mm wavelength presented
here.

For the polarization signal at mm wavelengths to be produced by
scattering off dust particles, the presence of grains with $a_{\rm
  grain} \sim \lambda/2\pi$ is required (for our observations this
means $a_{\rm grain} \gtrsim 250~\mu$m). This is because the
scattering cross-sections are too small unless the grains are made
large. However, if the grains are too large, $a_{\rm grain} \gtrsim
500 \mu$m for observations at 1 mm, the polarization efficiency
becomes too small and polarized light is not efficiently produced
\citep[see, e.g.][]{Kataoka2015}. Scattering of radiation from a point
source due to randomly aligned dust particles in an optically thin medium
produces polarization vectors tangential to the radiation
field. Hence, in this scenario, our observations of non-tangential
polarization vectors would imply that the source of millimeter
radiation is more complex, that the grains have a preferable alignment
direction, or that the photons suffer multiple scattering episodes.
We disfavour multiple scattering as a possible explanation because the
effect of high scattering optical depths is to decrease the
polarization degree, as shown by \citet{Kataoka2015}. Their models
show that a maximum polarization degree of $2.5\%$ is obtained in the
optically-thin regime and decreases for larger optical depths.
Regarding the asymmetry of the incident radiation field, the source of
millimeter flux is clearly not point-like, with detectable emission
arising from the star itself and from an extended N-NE feature, apart
from clump C. The direction of the polarization vectors cannot be
explained by the N-NE feature contributing to the source of
unpolarized photons, because that would cause the vectors to be
rotated in the opposite direction we observe. If we consider photons
emitted by clump C itself to be the primary source of unpolarized
radiation, the resolution of the observations would cause any
polarization created to be erased, unless Clump C has a very special
morphology. We consider this explanation unlikely, as higher
angular-resolution observations indicate that Clump C is elongated but still
approximately spherically symmetric (OG+15). Moreover, none of
these possible solutions is able to explain the direction of the
polarization vectors in visible wavelengths naturally.

The polarization at millimeter wavelengths can also be explained by
emission of dust particles aligned with the local magnetic field.
 Several alignment mechanisms have been proposed \citep[For recent reviews,
 see e.g.][]{Voshchinnikov2012, Andersson2015}. The
currently favoured theory for this alignment is the Radiative
Alignment Torque (RAT) theory \citep[e.g.][]{Lazarian2007}. In this
case, the grains are expected to spin about their minor axis, with
their major axis aligned perpendicularly to the magnetic field
lines. Hence, the direction of polarization of photons emitted by the
dust would be perpendicular to that of the magnetic field lines.  This
scenario can also explain the direction of the polarization seen in
visible wavelengths, because absorption and scattering rather than
emission are expected to dominate at visible wavelengths. Therefore,
background scattered visible light would be more efficiently absorbed
or scattered out of the line-of-sight in the direction of grain
alignment. The resulting polarization would then have a direction
parallel to that of the magnetic field lines and perpendicular to the
polarization observed in emission. This is exactly what we find in our
observations.

The RAT theory was mainly developed for interstellar dust grains. To
investigate if the mechanism would work in the much denser
circumstellar environment we need to compare the alignment timescale,
given by the Larmor precession timescale, with the timescale set by
gas-dust interaction \citep[e.g.][]{Hoang2009, Reissl2016}. This
comparison yields a lower limit for the magnetic field strength that,
following \citet{Hoang2009}, can be written as:
\begin{equation}
|B|>4.1\times10^{-11} {{a}\over{s^2}} n_g T_d \sqrt{T_g} ~ {\rm [G]}.
\end{equation} 
We assume typical grains with a grain size of
$a=10^{-5}$~cm~$[=0.1~\mu$m$]$ and aspect ratio $s=0.5$. We now assume
the gas and dust temperature to be equal to the maximum allowed by the
observations of clump C, $T_d=T_g=100$~K. The gas particle density
$n_g$ is more uncertain, as no molecular gas has yet been detected in
clump C \citep[e.g. OG+15,][]{DeBeck2015}. Taking the dust mass
derived from our measured fluxes, the size of clump C from OG+15, and
assuming a gas-to-dust ratio of $100$, we find $n_g\approx
9\times10^9$~cm$^{-3}$. For the dust alignment to be efficient, we
find $B>13$~mG (or less if clump C is indeed severely depleted in
gas). This increases to $\sim65$~mG for grain sizes of $0.5~\mu$m that
are suggested to exist in part of the VY~CMa outflow
\citep{Scicluna2015}.  Even taking into account uncertainties on grain
size and aspect ratios, and gas temperature and density, these values
are still less than the field strength ($>100$~mG) derived from H$_2$O
masers at a similar distance to the star as clump C
\citep{Vlemmings2002}. Even in the strong radiation field of VY~CMa,
the radiative alignment timescale, following \citet{Hoang2009},
becomes of the order of a few years. This is still significantly more
than the Larmor precession timescale for a $13$~mG field of only
$\sim0.5$~day. This means that the grains will have had the time to
align with the magnetic fields.  Emission from aligned grains  under the RAT alignment theory can thus explain our observations.

In addition to the RAT alignment theory, a number of other grain
  alignment theories exist. Several of these theories make specific
  predictions and so far only the RAT theory can explain all
  observational characteristics of aligned dust
  \citep[e.g.][]{Voshchinnikov2012, Andersson2015}. However, different
  mechanism can operate or dominate in different astrophysical
  regions, and the special conditions in the dense circumstellar
  environment can provide further tests for the different alignment
  mechanisms. The earliest proposed mechanism is the imperfect
  Davis-Greenstein (IDG) mechanism \citep{DG51} that has seen several
  adjustments over the last decades. In this mechanism the dissipation
  of magnetization energy (for grains including small amounts of iron)
  aligns the spin axis with the magnetic field. This theory
  specifically aligns the small grains and the alignment efficiency
  depends on the temperature difference between the dust and the
  gas. Under our earlier assumptions, of equal dust- and
  gas-temperatures, we would thus not expect any alignment. However,
  further detailed observations of clump C are needed to determine if
  conditions are such that the IDG mechanism can operate. Thus, while
  our observations are fully consistent with the RAT theory, and
  provide strong constraints on the alignment timescales, we cannot
  yet rule out other mechanisms that align the grains with the
  magnetic field.

\subsection{Magnetic field morphology}

Having determined that the linear polarization of the majority of the
SiO masers traces the magnetic field, we notice, with the exception of
the potentially saturated maser feature at $V_{\rm
  lsr}=15$~km~s$^{-1}$, a clear structure in the $v=1$ $^{28}$SiO
EVPAs. The vectors rotate from $\sim130^\circ$ at
blue-shifted velocities to $\sim-50^\circ$ around the stellar velocity
and back to $\sim140^\circ$ on the red-shifted side. Almost exacly the
same rotating behaviour was seen in the 86~GHz $^{28}$SiO $J=2-1, v=1$
polarization vectors observed by \citet{Herpin2006}. The EVPAs we observe are also consistent with the angles measured for the
$^{28}$SiO ($J=5-4, v=1$) masers in \citet{Shinnaga2004}.

The actual magnetic field direction is parallel or perpendicular to
the measured polarization vectors, depending on whether the angle
$\theta$ between the maser propagation and the line of sight is
respectively smaller or larger than the Van Vleck angle
$\theta_{cr}\approx55^\circ$, where the fractional polarization is
equal to zero. Considering the dust polarization
vectors arising in clump C, which is thought to be close to the plane
of the stellar velocity, are perpendicular to the magnetic field
direction, the same appears to be true for the SiO masers. For most
masers, we can thus conclude that $\theta>55^\circ$ and that the
magnetic field is perpendicular to the SiO maser EVPA. That is also consistent with the $\sim40^\circ$ magnetc field
angle determined from the extreme red-shifted thermal SiO.

Such polarization behaviour may be consistent with a predominantly
toroidal field morphology, with an inclination $>55^\circ$ and a
position angle on the sky of $\sim40^\circ$. The magnetic field
strength dependence on radius of $B\propto r^{-1}$ for a toroidal
field also fits the field strength derived from the SiO masers
observed here and previously observations \citep{Herpin2006,
  Richter2016}, together with the OH, and H$_2$O masers
\citep{Vlemmings2002}.

The hypothesis of a predominantly toroidal magnetic field, however,
can only be confirmed with high angular resolution observations, as it
depends on the position of the individual maser components. Component
fitting of the SiO masers presented here indicates that they occur
within $\sim20$~mas, or $\sim2$~R$_*$, from the star. As the
positional uncertainty for most masers is of similar order, it is not
possible to produce a more detailed map of the magnetic field with the
current data.

Assuming a toroidal field configuration would place clump C in the
magnetic equator, possibly the result of a magnetically controlled
ejecta. This scenario is supported by the observations of
\citet{Richter2016}. Although on much smaller scales, their VLBI
observations indicate elongation of $v=1 J=1-0$ SiO maser features
from the star in the general direction of clump C. For these masers
they derive a $2-5$ Gauss magnetic field strength. \citet{Richter2016}
speculate that the elongated features, at least their "F1" feature
studied in detail, may be associated with a highly magnetized outflow
above a stellar surface magnetic cool spot or convective cell.

\section{Conclusions}

The ALMA Band 5 178~GHz continuum observations of VY~CMa reveal that
the dust structure at this wavelength is similar to the structure
found in previous 321, and 658~GHz ALMA observations. The derived
spectral index map indicates the dust of clump C is mostly optically
thick even at 178~GHz. Assuming standard dust properties, this means
the estimated clump C dust mass is $>1.2\times10^{-3}$~M$_\odot$.

Significant linear polarization of up to $3\%$ is detected from clump
C. This is the first detection of dust polarization at millimeter
wavelengths in the inner circumstellar environment of an evolved
star. The EVPA is $\sim8^\circ$ and is neither radial
nor tangential to the stellar radiation field.  If the continuum
polarization is the result of emission from magnetically aligned dust
grains, then the observations indicate that, according to RAT grain
alignment theory, the grains cannot be diamagnetic
(e.g. purely carbonaceous). This is not unexpected for the oxygen rich
environment of VY~CMa where for example Si, Mg, and/or Fe are expected
to be present in the dust. The magnetic field would have a preferred
angle of $98^\circ$ and should be $>13$~mG in strength.  As higher
field strengths of $0.1$ to a few Gauss are measured in the maser
region, the magnetic field thus has no problem aligning the dust
grains in clump C. Higher angular resolution observations could
subsequently be done to determine the magnetic field structure and its
possible role in confining the dense clump C. Although we show it to
be unlikely, we cannot yet completely rule out a self-scattering
origin of the polarization. In this case, the grains in clump C would
need to be extremely large $(>250~\mu$m$)$. The origin of the
preferred polarization direction would then also imply a very specific
dust geometry.

The linear polarization of the SiO masers is typically $\sim5-10\%$,
and goes up to $\sim20-30\%$ for specific features. The average
polarization is at the level that could be expected from regular maser
polarization theory of non- to intermediately saturated masers, and
does not require anisotropic pumping of the magnetic substates. The
EVPAis also neither tangential nor radial
and indicates a possible toroidal magnetic field morphology. For the
thermal SiO, the linear polarization appears consistent with a
direction set by the stellar radiation field at the blue-shifted side,
but not at the red-shifted side, where it also seems to probe the
magnetic field.

Circular polarization is also detected for the maser and thermal
lines. The magnitude and spectral bevaviour of the circular
polarization indicates it is unlikely to be caused by standard
instrumental effects or calibration errors. The observed polarization
is also consistent with previous measurements. Finally, since the
calibrators and VY~CMa continuum show no sign of circular polarization
and the Stokes V charactistics are very different for the maser and
thermal lines, we suggest that the circular polarization is likely
intrinsic to the source. However, as this constitutes the first
circular polarization imaged with ALMA, additional observations are
needed to rule out possible unknown instrumental contributions to
Stokes V and confirm our observations. A normal Zeeman interpretation
of the maser lines would imply a field of $\sim1-3$~G in the SiO maser
region of VY~CMa. This is consistent with previous lower frequency SiO
and H$_2$O maser observations and the toroidal field
interpretation. The circular polarization of the thermal line could
potentially originate from resonant scattering of magnetized SiO gas
at larger distance from the star.

\begin{acknowledgements}
  This work was supported by ERC consolidator grant 614264. WV and TK
  also acknowledge support from the Swedish Research Council. This
  paper makes use of the following ALMA data:
  ADS/JAO.ALMA\#2011.0.00011.S. ALMA is a partnership of ESO
  (representing its member states), NSF (USA) and NINS (Japan),
  together with NRC (Canada), NSC and ASIAA (Taiwan), and KASI
  (Republic of Korea), in cooperation with the Republic of Chile. The
  Joint ALMA Observatory is operated by ESO, AUI/NRAO and NAOJ. We
  thank the ALMA Band 5 Science Verification team for their efforts in
  preparing and performing the observations and the ARC node network
  reduction team and the hosts of the Nordic ARC for their reduction
  work.
\end{acknowledgements}

%
%


\begin{appendix}

\section{Monte Carlo analysis of spurious circular polarization}
\label{AppMC}

\begin{figure*}
\includegraphics[width=14cm]{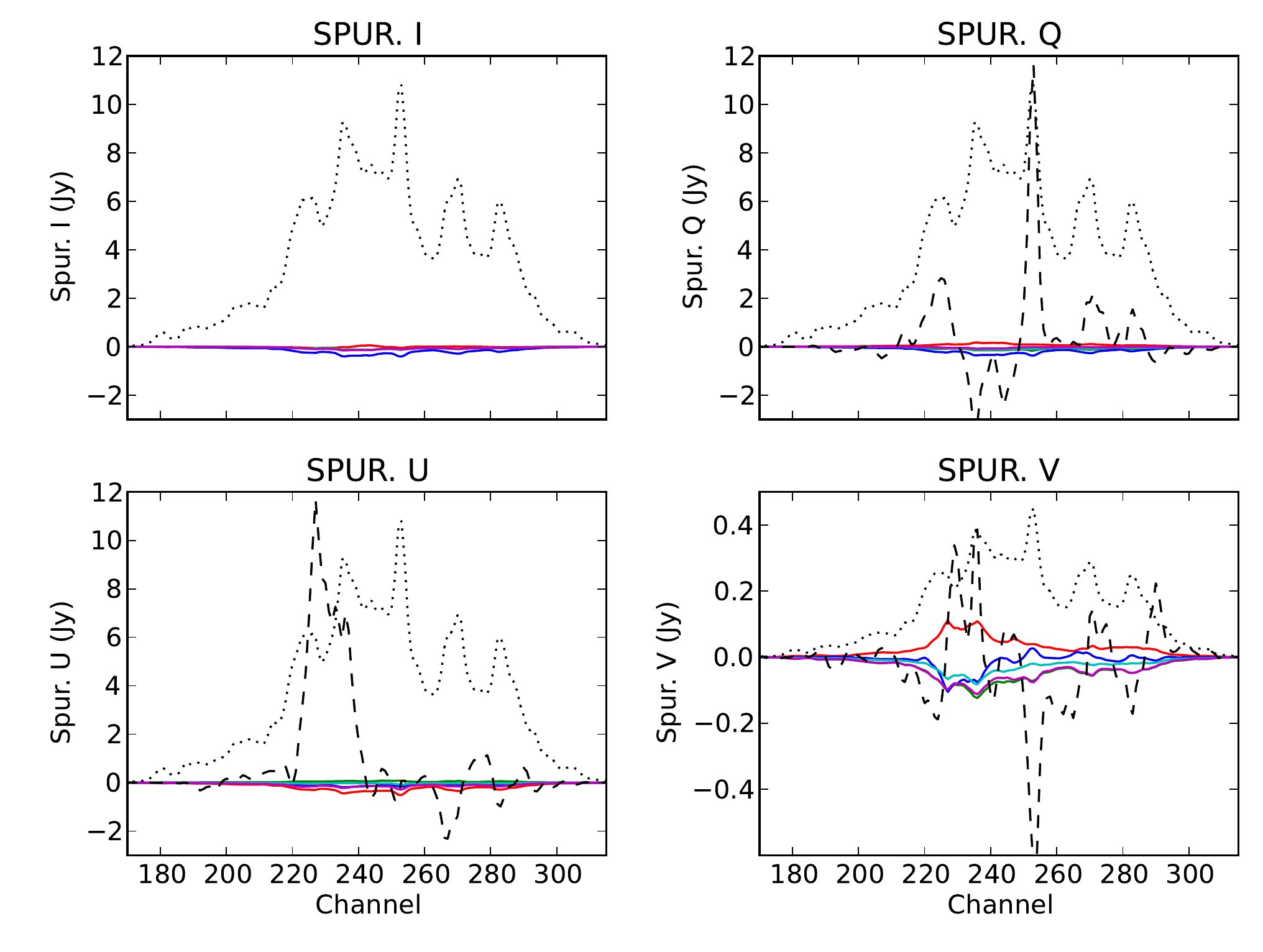}
\caption{Spurious polarization in spw\,3 (solid lines) for 5
  representative Monte Carlo simulations. The spectra are extracted at
  the position of the line peak. The quantities are given in Jy. The
  short dashed line indicates a total intensity spectrum (not to
  scale) for illustration. The long-dashed line are the observed Q, U,
  and V polarizations at the same scale as the simulated spurious
  polarization spectra. It is clear that the magnitude and spectral
  behaviour of the observed spectra is significantly different from
  the spurious signal.}
\label{MC_SPEC1}
\end{figure*}

\begin{figure*}
\includegraphics[width=14cm]{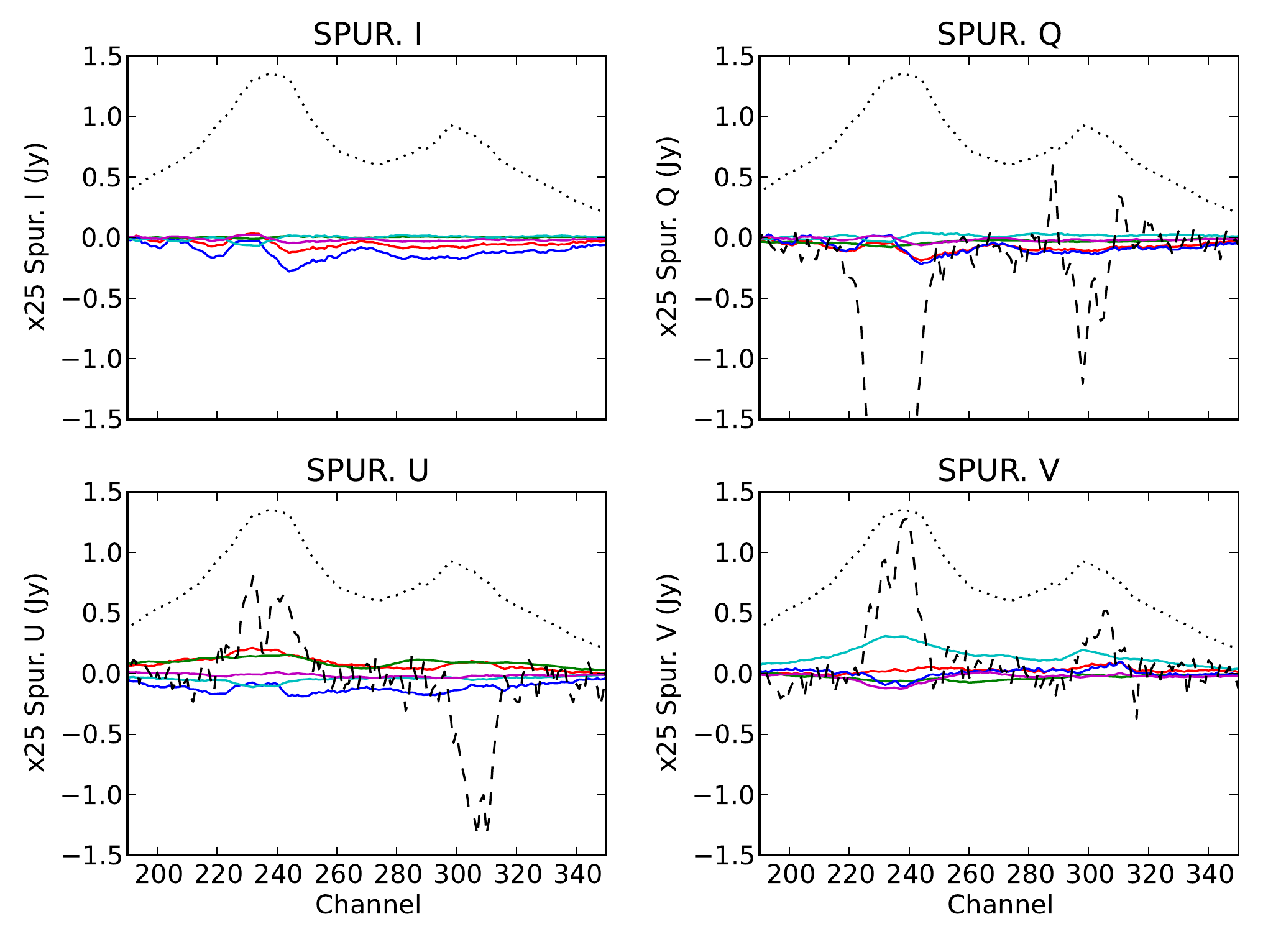}
\caption{Same as Fig.\,\ref{MC_SPEC1}, but for spw\,4. Note that the
  spectra are multiplied by 25.}
\label{MC_SPEC2}
\end{figure*}
 
The calibration of the instrumental polarization can be described by
two different quantities: on the one hand, the relative phase between
the polarizers at the reference antenna (we call this quantity {\em
  X-Y phase}) and, on the other hand, the leakage between the
polarizers of all the antennas, which are modelled by complex-valued
quantities called {\em D-terms}.  In order to simulate the effects
that a residual X-Y phase and/or polarization leakage would have on
the final SiO images in Stokes V, we have performed a Monte Carlo
analysis. We have assumed that both the X-Y phase and the D-terms are
smooth functions of the frequency within the band, so that the channel
noise in the calibration plots shown in \citep{IMV2016} reflects the
uncertainties related to the process of antenna-gain estimation. All
these uncertainties, combined, may bias the final images, introducing
spurious signals in all four Stokes parameters.

Judging from the calibration plots in \citep{IMV2016} (their Fig.\,8),
we estimate an upper bound in the X-Y phase uncertainty of $\sim$10
degrees. Regarding the D-terms (Fig.\,11), the typical scatter among
neighbouring frequency channels (of the same antenna) is 1$-$2\%.

Based on this assumptions, we have generated one thousand sets of
random D-terms for all the ALMA antennas, together with random X-Y
phases for the reference antenna. Then, we applied these quantities to
the calibrated data and generated modified images of VY\,CMa for all
four Stokes parameters.  The random D-terms in our analysis follow a
Gaussian distribution with a 1$\sigma$ amplitude of 1\%. The X-Y phase
follows a Gaussian distribution with a 1$\sigma$ deviation of
10\,degrees.
 
After the imaging, we have computed the {\em difference} between the
calibration-corrupted full-polarization images and those obtained from
the original calibration (we call this difference the {\em spurious
  source polarization}). In Figs. \ref{MC_SPEC1}, and \ref{MC_SPEC2},
we show the spurious polarization (in all four Stokes parameters)
resulting from 5 representative Monte Carlo iterations for spw\,3, and
4. The spectra are taken at the peak intensity of Stokes I.

A seen in Fig.~\ref{MC_SPEC1}, we notice that the spurious fractional
Stokes V has typical wide-band average values of the order of
$\pm$0.1\% across the entire SiO line. This is similar to the {\em
  measured} average Stokes V from the original polarization
calibration: $-0.1$\% and indicates our assumed uncertainties in the
calibration are correct.

In any case, the spurious Stokes V in our Monte Carlo analysis is
typically below 0.1\% in absolute value, and has a smooth frequency
dependence across the line. On the contrary, the Stokes V recovered
from the original calibration shows a much stronger frequency
dependence, with extreme changes between close by channels that are up
to 4--5 time higher than those seen in the spurious Stokes V. This is
a clear indication that the {\em differential} Stokes V (i.e., the
frequency dependence of Stokes V) is a robust quantity, almost
independent of calibration biases. A similar conclusion can be derived
from the Monte Carlo analysis on spw\,4 (Fig.\,\ref{MC_SPEC2}). We can
thus conclude that the observed Stokes V behaviour of the SiO lines is
likely real and not an artifact of calibration. However, to rule out
an unkown instrumental effect introduced in the ALMA system would
require further observations.

\end{appendix}

\end{document}